Published version:
> Fashami, Mohammad Salehi, et al. "Switching of dipole coupled multiferroic nanomagnets in the presence of thermal noise: Reliability of nanomagnetic logic." *IEEE Transactions on Nanotechnology* 12.6 (2013): 1206-1212.

**DOI:** 10.1109/TNANO.2013.2284777

# Corrigendum to "Switching of Dipole Coupled Multiferroic Nanomagnets in the Presence of Thermal Noise: Reliability of Nanomagnetic Logic"

Mohammad Salehi Fashami, Kamaram Munira, Supriyo Bandyopadhyay, *Fellow, IEEE*, Avik W. Ghosh, *Senior Member, IEEE,* and Jayasimha Atulasimha, *Senior Member, IEEE*

The following typographical errors are present in the above paper [1]:

1. Equations 1, 5 and 6, as well as the text in Section III. C (Page 1210): The demagnetizing factors in the x, y and z directions should have been printed respectively as: $N_{d\_xx}$ $N_{d\_yy}$ and $N_{d\_zz}$. The letters denoting the directions are correct in all instances but the use of subscripts/hyphens have been mixed/omitted in some instances.

2. Equations 2: The random magnetic field $H_{thermal}$ due to thermal noise wrongly appears as $H_{thermal}$.

3. On page 1208, left column, 4th row from the bottom: (Fig. 1.b) is wrongly referred to as (. 1.b).

Furthermore, our stochastic simulations had a systematic error that lead to our reporting a higher error than the correct estimate in Figures 2 and 4 of the original paper [1]. The corrected versions of these figures are Figure 1 and Figure 2, respectively, in this corrigendum. However, the new plots convey the same physics and overall message: dipole coupled architectures are unlikely to meet the stringent error requirements of conventional Boolean logic. Nonetheless, such architectures may have niche applications where energy dissipation is the main consideration and accuracy is not. One example is implanted medical devices that cannot drain the battery too quickly, but need not be too accurate since they usually diagnose a patient's condition by approximately matching signals.

Hence, the key message and our conclusions remain unchanged.

The only change is in the values of the error quoted. With the new error values, the following sentences would be modified:

a. Page 1209, last but one line: "However, even with slow stress withdrawal (~5 ns), moderate dipole coupling (~200 nm pitch), and stress (4 MPa) close to critical stress (3.145 MPa), the error rate is 3.93% as shown in Fig 2 of this corrigendum.

b. Page 1210, paragraph just before CONCLUSION: "Nonetheless, to verify if higher dipole coupling alleviates the problem and reduces error rate, we performed a simulation with inter-magnet spacing of 120 nm (the minimum allowed so the ground state remains anti-ferromagnetic) and the same ~5 ns stress withdrawal time. The error probability was ~0.001% (1 in 100,000) in this case, which is still very high for traditional Boolean logic."

### Acknowledgements

We thank Mr. Mamun Al Rashid for benchmarking the results in Figure 1 and 2 of this corrigendum.

M. Salehi Fashami is with the Department of Physics, University of Delaware, Newark, DE 19716 USA (e-mail: mfashami@udel.edu).
K. Munira is with the Center for Materials for Information Technology, University of Alabama, Alabama 35405 USA (e-mail: kmunira@mint.ua.edu).
S. Bandyopadhyay is with the Department of Electrical Engineering, Virginia Commonwealth University, Richmond, VA 23284 USA (e-mail: sbandy@vcu.edu).
A. W. Ghosh is with the Department of Electrical and Computer Engineering, University of Virginia, Charlottesville, VA 22903 USA (e-mail: ag7rq@virginia.edu).
J. Atulasimha is with the Department of Mechanical and Nuclear Engineering, Virginia Commonwealth University, Richmond, VA. 23284 USA (e-mail: jatulasimha@vcu.edu).

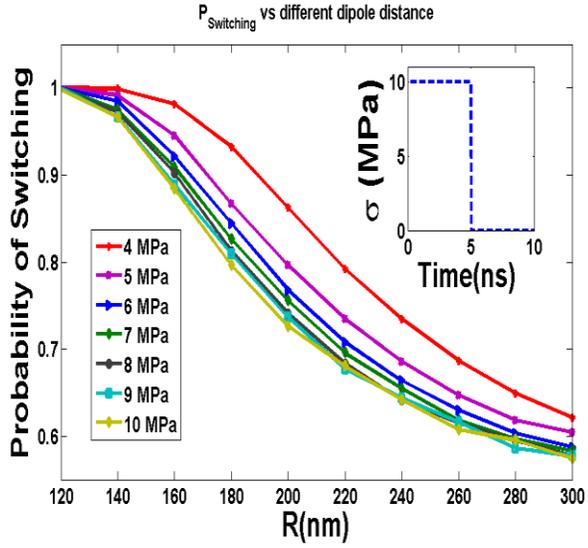 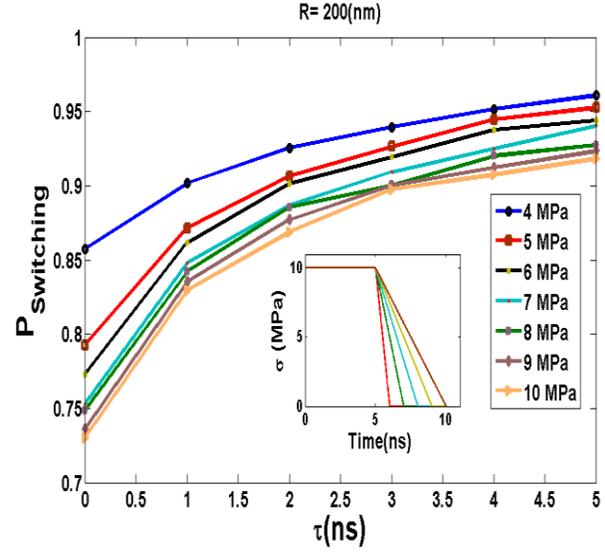

**Fig. 1:** Probability of stress-induced switching of the right nanomagnet in response to the magnetization state of the left versus separation between the two nanomagnets (R). Plots are presented for different stress values assuming that stress is turned on and off abruptly. Right nanomagnet dimension: 105 × 95 × 5.8 nm. INSET: compressive stress is applied for 5 ns and released.

**Fig. 2:** Probability of switching versus time for raising the energy barrier, i.e., time for withdrawal of stress, at fixed pitch R = 200 nm and different stress levels. Nanomagnet dimension: 105 × 95 × 5.8 nm. Stress withdrawal in ~ 1 ns is reasonable as the piezoelectric response is typically ~100 ps [2]. INSET: compressive stress is applied for 5 ns and released in different linear ramps varying in duration from 1 to 5 ns.

# Switching of Dipole Coupled Multiferroic Nanomagnets in the Presence of Thermal Noise: Reliability of Nanomagnetic Logic

Mohammad Salehi Fashami, Kamaram Munira, Supriyo Bandyopadhyay, Fellow, IEEE,

Avik W. Ghosh, Senior Member, IEEE and Jayasimha Atulasimha*, Senior Member, IEEE

*Abstract*—The stress-induced switching behavior of a multiferroic nanomagnet, dipole coupled to a hard nanomagnet, is numerically studied by solving the stochastic Landau-Lifshitz - Gilbert (LLG) equation for a single domain macro-spin state. Different factors were found to affect the switching probability in the presence of thermal noise at room temperature: (i) dipole coupling strength, (ii) stress levels, and (iii) stress withdrawal rates (ramp rates). We report that the thermal broadening of the magnetization distribution causes large switching error rates. This could render nanomagnetic logic schemes that rely on dipole coupling to perform Boolean logic operations impractical whether they are clocked by stress or field or other means.

*Index Terms*- Landau-Lifshitz-Gilbert equation (LLG equation) Nanomagnetic logic, reliability, thermal noise.

## I. INTRODUCTION

Dipole coupled shape-anisotropic nanomagnets with bistable magnetization are a popular platform for implementing Boolean logic circuits [1-2]. Because single domain nanomagnets flip by coherent spin rotation [3-4], they should dissipate very little energy (< 1 aJ) to switch, which ought to make nanomagnetic logic (NML) far more energy-efficient than traditional transistor-based logic. Unfortunately, Bennett clocking is required to steer bits unidirectionally from one stage to another in dipole coupled arrays [5] and energy-inefficient Bennett clocking schemes may offset any possible energy advantage that magnets have.

We have devised an extremely energy efficient Bennett clocking scheme employing strain-induced switching of *multiferroic* nanomagnets which may result in superior nanomagnetic logic [6-11] and memory [12-15] that consume very little energy to operate. This paradigm has been termed "hybrid straintronic-spintronic nanomagnetic logic (SNL)" since the magnetization of a nanomagnet is rotated with electrically generated mechanical strain.

The basic bistable switch in SNL is a two-phase multiferroic nanomagnet consisting of a piezoelectric layer elastically coupled to a magnetostrictive layer (Fig. 1 a).

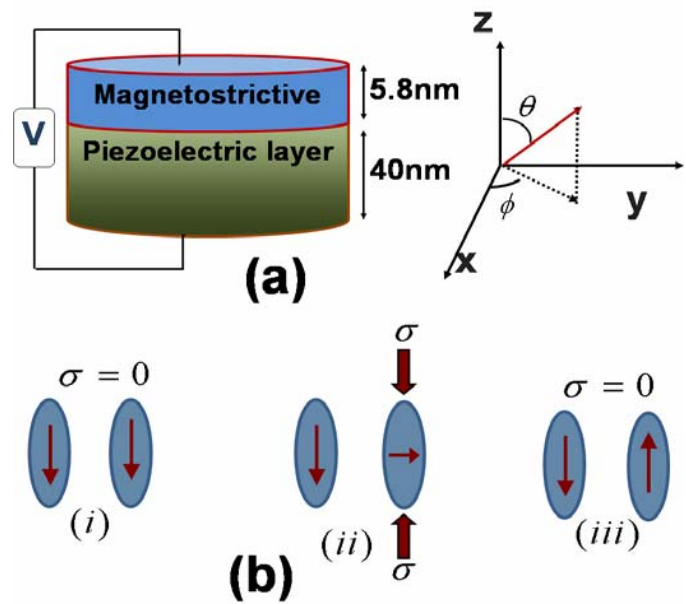

Fig. 1: (a) An elliptical multiferroic nanomagnet consisting of a piezoelectric layer in intimate contact with a magnetostrictive layer.
(b) A dipole-coupled nanomagnet system comprising a hard nanomagnet (left) with large shape anisotropy and a soft multiferroic nanomagnet (right) with smaller shape anisotropy whose shape anisotropy energy barrier is modulated with stress. (i) The magnetizations are rendered parallel by aligning the left magnet's magnetization along that of the right magnet's magnetization with an external agent. This does not immediately flip the right magnet since the dipole coupling is not strong enough to overcome the latter's shape anisotropy energy barrier. (ii) Upon application of sufficient compressive stress to the right nanomagnet, the shape anisotropy barrier is inverted and the magnetization rotates close to the hard axis. (iii) Upon removal of stress, the magnetization of the second magnet settles to a state anti-parallel to the first since this configuration becomes the global energy minimum.

A voltage pulse applied across the piezoelectric layer generates a strain pulse which is transferred to the magnetostrictive layer and flips its magnetization because of

This work is supported by the U.S. National Science Foundation under the SHF-Small grant CCF-1216614, NEB 2020 grant ECCS-1124714 and by the Semiconductor Research Corporation (SRC) under NRI Task 2203.001.

M. Salehi Fashami and J. Atulasimha are with the Department of Mechanical and Nuclear Engineering, Virginia Commonwealth University, Richmond, VA 23220, USA. *Corresponding author jatulasimha@vcu.edu
S. Bandyopadhyay is with the Department of Electrical and Computer Engineering, Virginia Commonwealth University, Richmond, VA 23220, USA.
A.W. Ghosh and K. Munira are with Charles L. Brown Department of Electrical and Computer Engineering, University of Virginia, Charlottesville, VA 22903, USA.

the generated stress [12, 13]. In an isolated magnet acting as a memory element, simple magnetization flipping is all that is needed to write bits, but in a logic circuit that computes and produces an output depending on the inputs, conditional flipping based on the state of other magnets is required. This is accomplished by dipole coupling the output magnet to the input magnet(s). The output magnet switches its state as dictated by the input magnet states, after it receives the appropriate clock (voltage) pulse(s). There is limited reliability analysis for this process in the presence of thermal noise [16 - 18], but it is critical for assessing the viability of dipole coupled nanomagnetic logic.

In this paper, we systematically study the influence of dipole coupling strength, stress levels (clock amplitude), and stress withdrawal rates (clock ramp rate) on the switching probability of a dipole coupled nanomagnet pair shown in Fig 1 b, in the presence of thermal noise. We further report that dipole coupled nonmagnetic computing is likely to be too error-prone for Boolean logic. A more detailed study of error probability when "pulse shaping", (the use of different clock waveforms) is employed has been presented elsewhere [11].

## II. MODELING MAGNETIZATION DYNAMICS IN THE PRESENCE OF THERMAL NOISE

We consider a pair of shape-anisotropic nanomagnets spaced far enough apart that the ground state magnetizations of the two magnets are mutually anti-parallel. This system is shown in Fig. 1 b.

It is assumed that the magnetization of the left nanomagnet is stiff while that of the right nanomagnet rotates under the influence of both the dipole field exerted by the left nanomagnet and the clock-induced stress. We note that the "stiffness" condition imposed on the left nanomagnet is only an artificial condition introduced here to avoid simulation of the entire array while still capturing the essential physics of switching via dipole coupling in the presence of thermal noise. Each nanomagnet has the shape of an elliptical cylinder as shown in Fig. 1 a, with a magnetostrictive layer ~5.8 nm thick deposited on a ~40nm thick PZT layer. This ensures that most of the strain generated in the PZT layer is transferred to the magnetostrictive layer through elastic coupling. It is further assumed that the PZT layer is mechanically constrained from expanding in the x-direction (see Fig. 1 a) so that it generates a uniaxial strain along the major axis (y-direction) through $d_{31}$ coupling when an electric field is applied across the PZT layer (in the z-direction). However, such clamping is not required if we use lateral contacts so that the $d_{33}$ coupling produces compressive strain along the major (in-plane easy) axis while the $d_{31}$ coupling produces tensile strain along the minor (in-plane hard) axis and vice versa. The nanomagnet dimensions are assumed to be 105 nm × 95 nm × 5.8 nm in all cases, so that the in-plane shape-anisotropy energy barrier between the two orientations along the easy axis (major axis of the ellipse) is ~ 0.75 eV (or ~ 32 kT at room temperature). This barrier prevents spontaneous switching of magnetization between the two stable orientations along the easy axis in the presence of thermal noise [6]. The magnet dimensions also ensure that the magnet has but a single ferromagnetic domain at equilibrium.

**TABLE I**
Material Parameters and Geometric Design for Terfenol-D

| Symbol | Quantity | Value |
|---|---|---|
| $(3/2)\lambda_s$ | magnetostriction | $9 \times 10^{-4}$ [24], [25] |
| $M_s$ | Saturation Magnetization | $0.8 \times 10^6$ A m$^{-1}$ |
| Y | Young's modulus | $8 \times 10^{10}$ Pa [26] |
| $\alpha$ | Damping factor | 0.1 [27] |
| $a \times b \times t$ | Dimension | 105nm×95nm×5.8nm (for Fig 2 plots) |

The first step is to derive the potential energy of the single-domain soft nanomagnet on the right with uniform magnetization $\vec{M}(t)$. A point dipole assumption is made to calculate the dipole coupling between the two single-domain magnets. The dipole coupling is likely to be overestimated in a macro-spin model where the magnetic dipole is assumed to act at a point (center of the magnet) compared to an OOMMF model that accounts for the spatial magnetization distribution in each nanomagnet. Nevertheless, this does not change the message of our paper, as we report that nanomagnetic logic is highly unreliable. If the dipole coupling is weaker than we estimate, the error rate would in fact be higher, i.e. nanomagnetic logic would be even less reliable than we predict.

The total energy of the soft (right) nanomagnet [6] is composed of the energy due to dipole coupling with the stiff left nanomagnet [19], shape-anisotropy [19] and stress-anisotropy caused by the clock pulse generating stress [19]:

$$U_i(t) = \underbrace{\frac{\mu_0 M_s^2 \Omega^2}{4\pi R^3} \left[ -2(\sin\theta_i \cos\phi_i)(\sin\theta_j \cos\phi_j) + (\sin\theta_i \sin\phi_i)(\sin\theta_j \sin\phi_j) + \cos\theta_i \cos\theta_j \right]}_{E_{dipole}}$$
$$+ \underbrace{\left(\frac{\mu_0}{2}\right) M_s^2 \Omega \left( N_{d\_xx} \left[\sin\theta_i(t)\cos\phi_i(t)\right]^2 + N_{d\_yy} \left[\sin\theta_i(t)\sin\phi_i(t)\right]^2 + N_{d\_zz} \left[\cos\theta_i(t)\right]^2 \right)}_{E_{shape-anisotropy}}$$
$$- \underbrace{\left(\frac{3}{2}\lambda_s \sigma_i \Omega\right) \sin^2\theta_i(t)\sin^2\phi_i(t)}_{E_{stress-anisotropy}} \quad (1)$$

The effective magnetic field $\vec{H}_{eff}^i(t)$ acting on the right nanomagnet at any instant of time $t$ is given by the gradient of the $U_i(t)$ with respect to its magnetization ($\vec{M}_i$).

$$\vec{H}_{eff}^i(t) = -\frac{1}{\mu_0 \Omega} \frac{\partial U_i(t)}{\partial \vec{M}_i(t)} + \vec{H}_{thermal} = -\frac{1}{\mu_0 M_s \Omega} \nabla_{\vec{m}} U_i(t) + \vec{H}_{thermal} \quad (2)$$

In equations (1) and (2), $M_S$ is the saturation magnetization of the magnetostrictive layer of the right nanomagnet; $\mu_0$ is the permeability of vacuum; $\gamma$ is the gyromagnetic ratio; $\Omega$ is the volume of the magnetostrictive layer; $\alpha$ is the Gilbert damping factor; $N_{d\_kk}$ is the demagnetization factor in the k$^{th}$ direction; $\lambda_s$ is the saturation magnetostriction and $R$ is the separation between two magnets.

All model parameters including material constants and geometric details are summarized in Table-1.

The effect of thermal fluctuation is modeled with a random field ($\vec{H}_{thermal}$) with statistical properties in the manner of ref [16, 20-21]. It is expressed as:

$$\vec{H}_{thermal} = \sqrt{\frac{2K_B T \alpha}{\mu_0 M_s \gamma \Omega \Delta t}} (\vec{G}(t)) \qquad (3)$$

where $\vec{G}(t)$ is a Gaussian random distribution with mean of 0 and variance of 1 in each Cartesian coordinate axis; $\Delta t$ is the time step used in simulating the switching trajectories and it is inversely proportional to the attempt frequency with which thermal noise disrupts magnetization. Finally, $K_B$ is the Boltzmann constant.

The effective magnetic field given by Equation (2), which represents the effect of dipole coupling, stress anisotropy, shape anisotropy as well as random thermal noise, is used in the vector stochastic Landau-Lifshitz-Gilbert equation [2, 3]

$$(1+\alpha^2)\frac{d\vec{M}(t)}{dt} = -\gamma \vec{M}(t) \times \vec{H}_{eff}(t) - \frac{\alpha \gamma}{M_s}\left[\vec{M}(t) \times \left(\vec{M}(t) \times \vec{H}_{eff}(t)\right)\right] \qquad (4)$$

to compute the temporal evaluation of magnetization vector $\vec{M}(t)$ of the right multiferroic nanomagnet under the simultaneous actions of the dipole interaction with its stiff left neighbor, its own shape anisotropy, stress and random thermal noise. Since the magnitude of the magnetization vector is invariant in time, we assume that in spherical coordinates, the orientation of this vector is completely described by the polar angle $\theta$ and azimuthal angle $\phi$, as shown in Fig. 1 a. We assume a single domain macro-spin approximation here, following Ref. [3], but non-equilibrium dynamics may produce some deviations from this assumption [22]. Therefore, we perform a OOMMF simulation in Appendix-1 to compare it with the macro-spin model and show that the macro-spin assumptions are likely to hold for the sample sizes considered.

We assume that the magnetization vector of the right nanomagnet is initially aligned along a stable state, i.e. one of two possible orientations along the easy (major) axis of the magnetostrictive layer. An external agent flips the left stiff magnet and puts the system in a metastable state where the magnetizations of the two magnets become temporarily parallel (. 1.b). The right magnet is then clocked to generate stress, which will attempt to kick the system out of the metastable state into the global ground state by flipping the magnetization and aligning it along the other stable orientation along the easy axis. If and when this happens, the two magnetizations become anti-parallel.

In order to simulate the different switching trajectories (under the random thermal field) by solving Equation (4), we picked the initial angle randomly from the Boltzmann distribution and ran 20,000 simulations. The simulation is terminated when $\phi$ approaches within $5^0$ of $\pm 90^0$. If switching terminates near $\phi_i = -90^0$ (initial orientation), then we conclude that switching *failed*, whereas if it terminates near $\phi_i = +90^0$, we conclude that switching *succeeded*. Based on this, we found the switching error probability (fraction of switching trajectories that ended in failure) as well as the time evolution of the distribution of magnetization orientations during withdrawal of stress. In general, we hold the stress long enough so the magnetization distribution reaches equilibrium before withdrawal of stress. The key question we seek to answer is how the strength of dipole coupling, stress magnitudes, stress withdrawal rates (clock ramp rate) affect switching error probability.

### III. ANALYSIS OF SWITCHING ERROR IN THE PRESENCE OF THERMAL NOISE

In this section, we analyze the influence of the following parameters on switching error probability:

A) Dipole coupling strength (varied by varying the spacing between the magnets) for different stress levels.

B) Stress withdrawal rate (ramp down time 1 ns to 5 ns) for different levels of stress at an inter magnet spacing of $R = 200$ nm.

C) Use of thick (~15 nm) and closely spaced nanomagnets to increase dipole coupling to see if the error can be decreased.

The total error probability is the fraction of times the magnetization of the soft right magnet fails to switch from parallel to anti-parallel configuration upon application of stress, or erroneously switches from anti-parallel to parallel configuration upon application of stress. We do not consider the latter possibility here since its dependence on these factors will not be any different (qualitatively) if we hold the stress until local equilibrium has been established.

#### A. EFFECT OF DIPOLE COUPLING AT VARIOUS STRESS LEVELS

The effect of dipole coupling on switching error at various stress levels is shown in Fig. 2. It should be noted that the stress levels studied here are all above the critical stress defined as the stress at which the stress anisotropy energy equals the shape anisotropy energy barrier. This is the minimum stress needed to overcome the shape anisotropy energy barrier and make it possible for the magnet to switch:

$$E_{Stress-anisotropy} = E_{Shape-anisotropy} \Rightarrow -\frac{3}{2}\lambda_S \sigma_{Critical}\Omega = [N_{d-xx} - N_{d-yy}](\mu_0/2)M_s^2\Omega \quad (4)$$

$$\sigma_{Critical} = \left|\frac{[N_{d-xx} - N_{d-yy}](\mu_0/2)M_s^2}{-\frac{3}{2}\lambda_S}\right| \sim 3.145 \text{ MPa (compressive)} \qquad (5)$$

In Fig. 2, the error probability decreases with increasing dipole coupling (smaller spacing R) independent of stress. Further, for any given dipole coupling strength, the error probability increases with increasing excess compressive stress (above the critical stress).

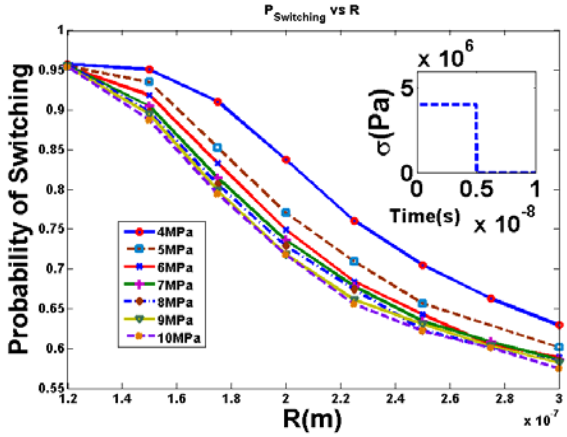

Fig. 2: Probability of switching vs. separation between the two nanomagnets (R) at different stress rates for sudden withdrawal of stress. Right nanomagnet dimension: 105×95×5.8 nm. INSET: Compressive stress is applied for 5 ns and released.

Both these trends can be explained using the schematic shown in Fig. 3. Clearly, when the dipole coupling strength is greater (spacing between nanomagnets is smaller), the energy landscape in Fig. 3 is such that the magnetization distribution is greatly skewed towards the anti-parallel state and hence the probability that it would end up in the parallel (wrong) state upon stress withdrawal is smaller. Now, for a given dipole coupling strength, excess stress makes the magnetization distribution less skewed towards the anti-parallel state, since the energy profile is modified by stress (Fig. 3) to increase the likelihood of the magnetization aligning close to the hard axis. This is why too much stress is undesirable and increases the error rate.

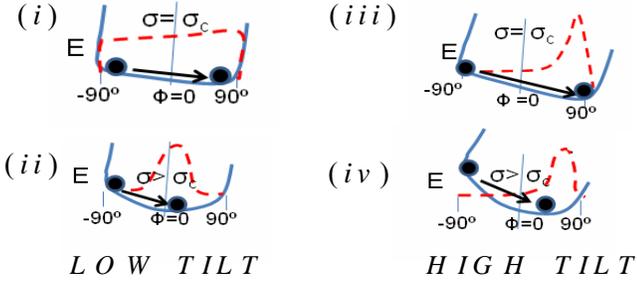

Fig. 3: Schematic that shows the effect of dipole coupling (tilt/asymmetry) and stress on the energy profile and the magnetization distribution. The distribution affects the dynamic switching and ultimately affects the dynamic switching error probability (which is smallest for critical stress and high dipole strength). The solid line shows the energy profile and the broken line depicts the corresponding probability distribution function for the magnetization orientation. (i) Low tilt (weak dipole coupling), critical stress ($\sigma = \sigma_C$), (ii) Low tilt (weak dipole coupling), high stress ($\sigma > \sigma_C$), (iii) High tilt (strong dipole coupling), critical stress ($\sigma = \sigma_C$), (iv) High tilt (strong dipole coupling, high stress ($\sigma > \sigma_C$).

### B. EFFECT OF STRESS WITHDRAWAL TIME (RAMP RATE) AT VARIOUS STRESS LEVELS

The effect of stress withdrawal rate is less intuitive. We analyze the effect of stress withdrawal rate (ramp time of 1 ns, 2 ns, 3 ns, 4 ns and 5 ns) in Fig. 4 for a large range of stresses for a fixed dipole coupling strength corresponding to a spacing of 200 nm. The trends clearly show that for all stresses, the error probability decreases with increasing stress withdrawal time (slower ramp is better). However, another important trend emerges: for fast stress withdrawal (~1ns): the error rate is strongly dependent on the stress level whereas for slower stress withdrawal rates (~5 ns), the error rate depends less critically on the applied stress. All the above trends can be explained using the schematic in Fig. 3. When the barrier (due to shape anisotropy) is suddenly raised (due to sudden stress withdrawal in ~1 ns) [see Fig. 1 (ii) and (iii) that describe the process, and Fig. 3, (i) and (ii) that show the in-plane magnetization distribution], the magnetizations that are skewed towards the anti-parallel state will switch correctly while those (minority) skewed towards the parallel state have insufficient time to correct themselves and therefore switch to the wrong state. Now, if the stress is withdrawn slowly (~5 ns), despite an initially high stress [unfavorable magnetization distribution, Fig. 3, (ii)] the energy profile has to gradually pass through the critical stress state [favorable orientation, Fig. 3, (i)] before the barrier is finally restored. In this case, even if a larger fraction of the magnetizations were originally skewed towards the parallel or wrong state [as in Fig. 3, (ii)], they have ample time to correct themselves and switch to the anti-parallel (correct) state as the energy profile gradually changes to favor switching to this state [as in Fig. 3, (i)]. However, even with slow stress withdrawal (~5 ns), moderate dipole coupling (~200 nm pitch) and stress (4MPa) close to critical stress (3.145 MPa), the error rate is ~ 5% as shown in Fig 4..This is obviously too high for logic applications.

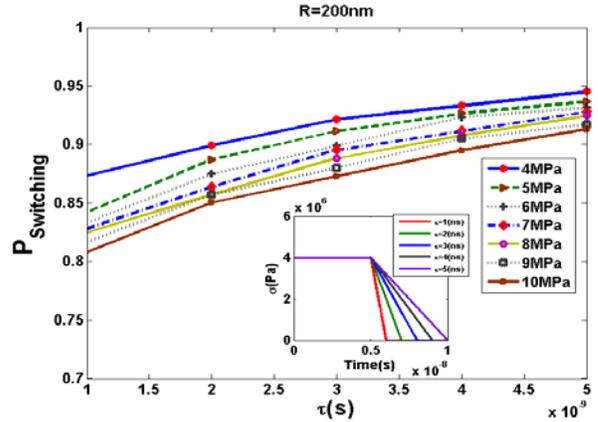

Fig. 4: Probability of switching vs. time for raising the energy barrier, i.e. time for withdrawal of stress, at fixed pitch =200nm and different stress levels. Nanomagnet dimension: 105×95×5.8 nm. Stress withdrawal in ~ 1 ns is reasonable as the piezoelectric response is typically ~100 ps [30]. INSET: Compressive stress is applied for 5 ns and released in different linear ramps varying in duration from 1 ns to 5 ns.

### C. INCREASED DIPOLE COUPLING STRENGTH

In our search to mitigate this problem, we examined whether it is possible to decrease the error probability to acceptable values by increasing the dipole coupling strength. This can be achieved by bringing magnets closer together, but too short a distance between them may make the dipole coupling energy larger than the shape anisotropy energy, at which point the bistability of the magnets will be lost and all magnets will couple ferromagnetically with their magnetization pointing along their hard axes. Thus, there is a minimum allowable spacing between magnets and the dipole coupling cannot be increased arbitrarily with this strategy. A second approach to increase dipole coupling strength is to use thicker nanomagnets to increase the magnet's volume without decreasing the magnet density (the dipole coupling strength is proportional to the square of the magnet's volume). If the thickness is increased to ~15 nm, the lowered $N_{d-zz}$ will necessitate using lateral dimensions ~102×104 nm for the ellipse to keep [$N_{d-xx}$ - $N_{d-yy}$] low enough that the in-plane shape anisotropy barrier remains at ~0.75 eV. This is necessary so that the shape anisotropy barrier can be easily overcome by stress anisotropy generated by the low "clocking" stress. Such precise geometric tolerance (2% difference between the major and minor axes of the ellipse) may not be feasible with current lithographic technology, but more importantly, decreasing the out-of-plane anisotropy results in a greater spread in the out-of-plane magnetization orientation during the switching process. A greater out-of-plane spread in magnetization orientation increases the error as shown elsewhere [28]. Therefore, this approach may be counter-productive.

Nonetheless, to verify if higher dipole coupling alleviates the problem and reduces error rate, we performed a simulation with inter-magnet spacing of 120 nm (the minimum allowed so the ground state remains anti-ferromagnetic) and the same ~ 5 ns stress withdrawal time. The error probability turned out to be ~ 1.75 % in this case, which is still very high. Therefore, it appears that there is no obvious route to making the error rate reasonable.

## IV. CONCLUSIONS

We have modeled the magnetization dynamics of a strain clocked multiferroic nanomagnet dipole coupled to a neighboring hard magnet in the presence of thermal noise. The purpose was to study *conditional* switching where the magnetization state of the soft magnet is conditioned on the state of the hard magnet, which is necessary for logic. We systematically study the effect of stress, dipole coupling and stress-withdrawal rate on the switching error and offer physical explanations as to why the error rate is minimized at slow stress withdrawal rates, high dipole coupling and intermediate stress magnitudes. However, even with the largest possible dipole coupling that would retain an anti-ferromagnetic ground state, the switching error probability is too large for Boolean logic. Therefore, dipole coupled logic appears to be innately unreliable.. The unreliability of dipole coupled nanomagnetic logic schemes has been highlighted in prior numerical [17] and recent experimental [23] work. Innovative pulse shaping schemes [11] and novel hardware error correction schemes may alleviate this problem to some extent, but it seems unlikely that the stringent error requirements of $10^{-15}$ error probability in Boolean logic can be met. This does not mean that "magnetic computing" is doomed; it merely points to a serious shortcoming of dipole coupled architectures. Boolean logic schemes that do not rely on dipole coupling [29] and non-Boolean computing schemes may still emerge as viable and energy efficient methods of computing with magnets.

## APPENDIX-1

In the main paper, we made the macro-spin approximation and modeled the magnetization dynamics in the nanomagnet as that of a single domain. Such a model allows study of the effects of thermal noise on magnetization dynamics via the stochastic LLG equation which would typically require ~20,000 Monte Carlo simulations. This would have been computationally intractable had we attempted a comparable number of micromagnetic simulations.

However, in order to show that the magnetization dynamics is likely to be coherent and uniform in such a nanomagnet system, we performed a three dimensional (3D) micromagnetic simulation using the Object Oriented Micro Magnetic Framework (OOMMF) simulation [31] and compared it with the macro-spin LLG model for the case where thermal noise is absent (i.e. the simulation is at 0K). Agreement between the two models will inspire confidence that the macro-spin model is realistic for our case.

The OOMMF performs time integration of Landau-Lifshitz-Gilbert (LLG) equation, where the effective field in each grid/cell is composed of the exchange, magnetostatic (shape anistropy) and external field terms.

$$\vec{H}_{eff} = \vec{H}_{exchange} + \vec{H}_{Shape\,anistropy} + \vec{H}_{external} \quad (A1)$$

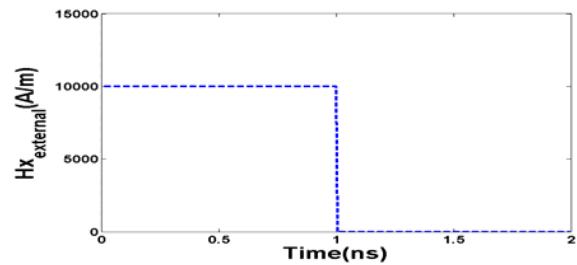

Fig.A1: External magnetic field in the X-direction applied to nanomagnet.

Since OOMMF does not have an explicit mechanism to incorporate the effect of stress, we can only model magnetic field induced magnetization dynamics and compare the macro-spin model and the OOMMF model for a sufficiently large field (~10kA/m) step (as shown in Fig. A1) applied along the positive x-axis. This external field completely erodes the shape anistropy barrier and makes the magnetization rotate through 90° from the magnetically easy direction to the magnetically hard direction. After 1 ns, the external field is removed and the barrier is restored. Thereafter, the magnetization rotates to the other easy direction under the influence of the dipole coupling from the neighboring nanomagnet which is modeled as a stiff magnet located at a distance of ~120 nm (center-to-center).

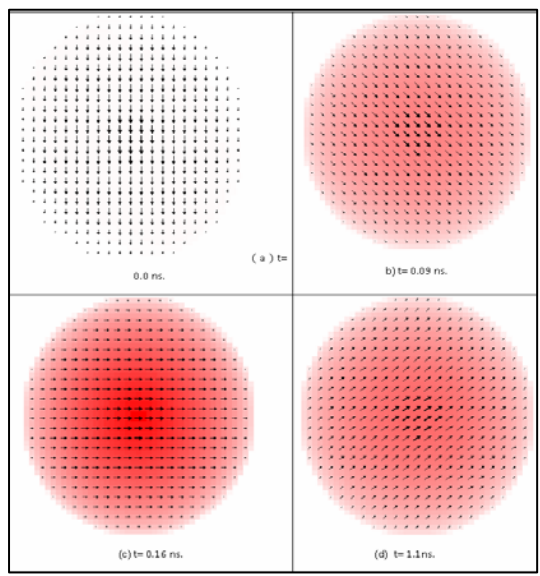

Fig. A2: An elliptical nanomagnet with size of 100nm×90nm×6nm is studied with OOMMF to explore if its magnetization rotates coherently with an external magnetic field of 10(kA/m).

In order to simulate the magnetization dynamics in a Terfenol-D magnetostrictive magnet with the OOMMF code, the following parameters were used in the simulations: saturation magnetization, $M_s = 800 \times 10^3 (A/m)$, exchange coupling stiffness term $A = 9 \times 10^{-12} (J/m)$ [32], anisotropy constant, $K_1 = 0 (J/m^3)$ (because we assume random polycrystalline orientation), and the Gilbert damping factor $\alpha = 0.1$. The cell size used for modeling was $2nm \times 2nm \times 2nm$. The dimension of nanomagnet was considered to be 100nm×90nm×6nm with elliptical lateral geometry as shown in Fig. A2.

The initial orientation of the magnetization vector is $\phi_0 = -90°$ and $\theta_0 = 89.9°$. Simulations were terminated when the residual torque fulfilled the condition: $|\vec{m} \times \vec{H}| < 10^{-6}$.

Fig. A2 shows that the magnetization dynamics predicted by OOMMF is close to that resulting from the macro-spin approximation and also the rotation of spins is mostly coherent, meaning that the magnetization rotates as a single domain. Furthermore, Fig. A3 shows that the average magnetizations in the x, y and z direction for the macro-spin and the LLG model agree within ~8%. The small disagreement is likely to be due to the small deviation from coherent rotation model during some phases of the switching process.

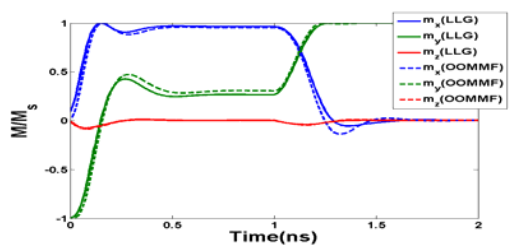

Fig. A3: Comparison of the OOMMF results with LLG simulation based on applying an external magnetic field.

ACKNOWLEDGMENT

We would like to acknowledge Dr. Michael Donahue at NIST for discussions on OOMMF simulations.

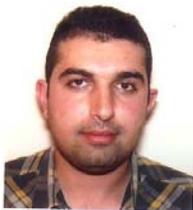
**Mohammad Salehi Fashami** received his B.Sc. degree in Physics from The Tarbiat Moallem University ,Tehran, Iran in 2004 and the M.Sc degree in Nuclear Reactor Engineering from The Amirkabir University of Technology (Tehran Polytechnic), Tehran, Iran in 2007. Since 2010, he has been working towards the Ph.D. degree in Mechanical Engineering at Virginia Commonwealth University, Richmond, VA, USA. His current research interests includes design, simulation and fabrication of strain clocked nanomagnetic logic devices using multiferroic nanomagnets and modeling of stochastic magnetization dynamics.

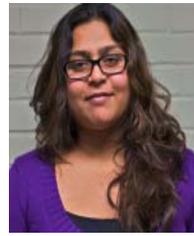
**Kamaram Munira** received the B.S. degree in computer science with a certificate in nanomaterials from Georgia Institute of Technology, GA, in 2006 and the Ph.D. degree in electrical engineering from the University of Virginia (UVa), Charlottesville, VA in 2012. She is currently a postdoctoral research fellow in Center for Materials for Information Technology at University of Alabama, Tuscaloosa, AL. Her current research areas include magnetic memories: Spin-Transfer Torque RAM (STT-RAM) and nanomagnetic logic. She was awarded the Louis T. Rader Graduate Research Award from the Department of Electrical Engineering at UVa in 2012.

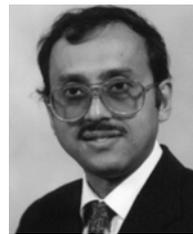
**Supriyo Bandyopadhyay** (SM'80–M'86–SM'88–F'05) is a Professor of Electrical and Computer engineering at Virginia Commonwealth University, Richmond, where he directs the Quantum Device Laboratory. He has authored and coauthored more than 300 scientific publications. His current research interests include nanoelectronics and spintronics. Prof. Bandyopadhyay chairs the Technical Committee on Spintronics within the IEEE Nanotechnology Council, and is a Fellow of the IEEE, the American Physical Society, the Electrochemical Society, the Institute of Physics and the American Association for the Advancement of Science.

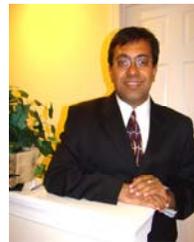
**Avik W. Ghosh** did his PhD in physics at the Ohio State University and postdoctoral research in Electrical Engineering at Purdue University. He is currently Associate Professor of Electrical and Computer Engineering at the University of Virginia. He is a Fellow of the Institute of Physics (IOP), senior member of the IEEE, and has received the 2011 IBM Faculty Award, the NSF CAREER Award for 2008-2013, and the Charles Brown New Faculty Teaching Award in 2006.

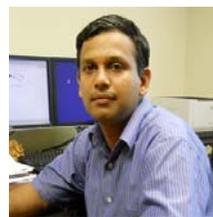
**Jayasimha Atulasimha** (SM'11) received the M.S. and Ph.D. degrees in Aerospace Engineering from the University of Maryland, College Park in 2003 and 2006, respectively. He is an Associate Professor of Mechanical and Nuclear Engineering at the Virginia Commonwealth University, Richmond, VA. He has authored or coauthored over 40 scientific articles including over 28 journal publications on magnetostrictive materials, magnetization dynamics and nanomagnetic computing. His research interests are magnetostrictive materials, nanoscale magnetization dynamics and multiferroic nanomagnet based computing architectures. He received the NSF CAREER Award for 2013-2018, and became a senior member of the IEEE in 2011.